\documentclass[11pt,letterpaper]{article}
\pdfoutput=1

\usepackage{apacite}

\usepackage{setspace}
\usepackage{latexsym,amsfonts,amsmath,amssymb}
\usepackage{amsthm}
\usepackage{mathtools} 
\usepackage{bbm}
\usepackage{graphicx}
\usepackage{color}
\usepackage{caption}
\usepackage{subcaption}
\usepackage[normalem]{ulem}
\usepackage{comment}
\usepackage{url}
\usepackage{slashed}
\usepackage{bm}
\usepackage{hhline}

\usepackage{tabu}

\newcommand{\CM}{\mathcal{M}}
\newcommand{\CN}{\mathcal{N}}

\newcommand{\CH}{\mathcal{H}}

\renewcommand{\Re}{{\rm Re}}
\newcommand{\Tr}{\mbox{Tr}}

\newcommand{\IR}{\mathbb{R}}
\newcommand{\IC}{\mathbb{C}}
\newcommand{\IZ}{\mathbb{Z}}

\newcommand{\IN}{\mathbb{N}}
\newcommand{\IH}{\mathbb{H}}

\newcommand{\rme}{e}
\newcommand{\ii}{\mathrm{i}}

\newcommand\be{\begin{equation}}
\newcommand\ee{\end{equation}}
\newcommand\bea{\begin{eqnarray}}
\newcommand\eea{\end{eqnarray}}

\renewcommand{\=}{\;= \;}

\renewcommand{\a}{\alpha}
\renewcommand{\b}{\beta}

\renewcommand{\t}{\tau}

\newcommand{\wt}{\widetilde}

\renewcommand{\Re}{\text{Re}}

\newcommand{\p}{\partial}

\newcommand{\ndt}{\noindent}

\newcommand{\half}{\frac12}

\renewcommand{\Re}{\mbox{Re}}

\title{Black holes and modular forms in string theory}

\author{Sameer Murthy}

\date{}

\begin{document}

\maketitle 
\begin{center}
{Department of Mathematics, King's College London, \\
Strand, London UK WC2R 2LS.}
\end{center}

\bigskip

\ndt {\bf Summary}

The study of black holes in string theory has led to the discovery of deep and surprising 
connections between black holes and modular forms---which are two classical, a priori unrelated, subjects.
This article explains the main physical and mathematical ideas behind these connections. 

It is known from the pioneering work of J.~Bekenstein and S.~Hawking in the 1970s that
black holes have thermodynamic entropy,
and should therefore be made up of a collection of microscopic quantum states. 
Superstring theory provides a framework wherein one can associate 
a number of microscopic states that make up the quantum-statistical system underlying a black hole, 
thus explaining their thermodynamic behavior from a more fundamental point of view. 
The basic connection to modular forms arises from the observation that, in the simplest superstring-theoretic construction, 
the generating function of the number of microscopic states is a modular form. 
In one direction, modular symmetry acts as a powerful guide to the calculation of quantum-gravitational 
effects on the black hole entropy. In the other direction, the connection has led to the discovery of surprising 
relations between Ramanujan's \emph{mock} modular forms and a class of string-theoretic black holes,  
thus providing an infinite number of new examples of mock modular forms. 

\bigskip

\ndt {\bf Keywords}

Black holes, black hole thermodynamics, string theory, modular forms, mock modular forms

\newpage


\section{Introduction}

\emph{Black holes} (BH) are objects that exist in the physical universe: they are regions of spacetime, typically formed by 
the collapse of heavy stars. At the theoretical level they are described
as solutions to the theory of General Relativity (GR). 
This theory is one of the pillars underlying the paradigm of modern physics---it determines all large-scale 
structure in our universe and underlies the dynamics of spacetime itself.
A black hole is characterized by the existence of a one-way surface surrounding it,  
called the \emph{event horizon}. The one-way nature of the horizon means that things can fall into the black hole but 
nothing---not even light---can come out, thus leading to its name. The theoretical description of black holes 
has been spectacularly confirmed by many recent observations in astronomy. 
See the textbook~\cite{Schutz:1985jx} for an introduction to GR and BHs.

\smallskip

\emph{Modular forms} are objects that exist in the mathematical universe: they are a class of functions 
that naturally arise in the field of number theory.  
They are characterized by their symmetry properties under the action of the \emph{modular group}~$SL(2,\IZ)$
(the group of~$2\times 2$ integer matrices with unit determinant).
The related term \emph{automorphic form} is used for 
functions with symmetry properties under more general arithmetic groups. 
Starting from the study of {theta functions} in the 19th century 
and its classic manifestations in number theory, the field of modular and automorphic forms has grown to 
include ramifications in topology, algebraic geometry, algebraic and analytic number theory, representation theory,
as well as physics, 
and has been the subject of intense study over the last 100 years. See~\cite{MF123} for an introduction to modular forms. 

\medskip

The work of~\cite{Bekenstein:1973ur} and~\cite{Hawking:1975vcx} 
showed that, when one takes into account quantum mechanical effects,  
black holes have thermodynamic entropy. This leads to the conclusion that a black hole must be made 
up of microscopic states, just like the entropy of air in a room is explained by the fact it consists of a large
number of molecules. 
The natural consequent question is:~\emph{What are the microstates (the ``molecules") of a black hole?}

Since no classical-mechanical process can probe the interior of a black hole,
the answer to the above question needs a framework which consistently takes into account 
quantum effects in the presence of a black hole.
Precisely such a framework is provided by superstring theory, wherein 
one can construct models of quantum black holes 
in which the number of microscopic states can be counted. 
The basic connection to modular forms arises from the observation that,  
in the simplest string-theoretic constructions, the number of these microscopic states 
is the Fourier coefficient of a certain modular form. 

The underlying modular symmetry allows us to make a simple asymptotic estimate for the logarithm of 
these Fourier coefficients, which agrees precisely with the Bekenstein-Hawking entropy of the corresponding BHs,
as was first discovered in the breakthrough work of~\cite{Strominger:1996sh}.
In fact, the power of modular symmetry allows us to go much further. 
The well-known Hardy-Ramanujan-Rademacher formula of analytic number theory, which expresses the Fourier
coefficient of the modular form as an infinite sum of~$I$-Bessel functions, 
turns out to be intimately related to quantum-gravitational corrections of the black hole entropy. 
The connection to black holes also informs the field of modular forms in mathematics. For example, 
it has led to new theorems and constructions of mock modular forms, 
(which were first introduced by Ramanujan in his last letter to Hardy in 1920).

\section{Black holes and their thermodynamics \label{sec:BHthermo}}

The structure and evolution of spacetime at large scales in our universe is described by the theory 
of General Relativity. 
Spacetime---the set of all points in space and time---is considered to be a pseudo-Riemannian 
4-manifold of signature~$(-,+,+,+)$, the~$-$~sign indicating the single time coordinate~$x^0$ and 
the three~$+$ signs related to the three spatial coordinates~$x^1,x^2,x^3$. 
The spacetime manifold comes equipped with a metric (a rank-2 tensor field with components~$g_{\mu\nu}(x)$ 
$\mu, \nu = 0,1,2,3$) which can be thought of, roughly, as encoding the local shape and size of the manifold. 
At large scales, physical matter in the universe is effectively described by 
a rank-2 tensor field on the spacetime manifold, called the stress-energy tensor with  
components~$T_{\mu\nu}(x)$.

The basic equations of general relativity, also referred to as the Einstein equations, are written 
in terms of differential geometric quantities, i.e.~the Ricci 
tensor field~$R_{\mu\nu}(g_{\mu\nu}(x))$ and its trace called the Ricci scalar field~$R(g_{\mu\nu}(x))$, 
which are measures of the curvature of the spacetime metric. 
The equations relate the spacetime curvature to the stress-energy tensor  
(for zero cosmological constant) as follows, 
\be \label{EinEqns}
R_{\mu \nu} - \frac12 g_{\mu \nu} R \= 8 \pi G \, T_{\mu \nu} \,,
\ee
where~$G$ is Newton's gravitational constant.

All physical structures at large scales in our universe, i.e.~galaxies, stars, etc, are solutions to the Einstein 
equations~\eqref{EinEqns}. 
Black holes are a very special class of solutions to these equations with the property that the 
resulting spacetime metric has a region from within which no classical signal can propagate out to an external observer. 
The boundary of this region is called the event horizon, the area of which is a measure of the size of the black hole. 
Black holes can have very complicated 
dynamics such as binary mergers\footnote{See e.g.~\url{https://www.ligo.org/}},   
but the solutions of Einstein's equations that describe stationary black holes 
are very simple.\footnote{Stationary means that they do not change as a function of time. One can think of these
as the end point of the complicated dynamical processes, when the black holes ``settle". This is the limit when
they can be studied as essentially independent objects.
}
These solutions are characterized by their conserved charges, namely their mass, spin, and 
electromagnetic 
charge.\footnote{Indeed, the dynamics of stationary charge-neutral black holes that are observed in the sky 
(see e.g.~\url{https://www.cfa.harvard.edu/research/topic/black-holes}) 
are well-described by the Kerr BH metric solution which is completely characterized by mass and spin.} 

\medskip

Theoretical investigations on black holes in the 1970s, primarily due  
to~\cite{Bekenstein:1973ur} and~\cite{Hawking:1975vcx}
established the following remarkable fact.
Although nothing can come out of a black hole in the classical theory of general relativity, a black hole 
in the quantum theory actually behaves like a thermodynamic object with associated temperature and entropy.

The main arguments to demonstrate the thermodynamic behavior of black holes are simple but profound, and 
proceed as follows. 
Consider a system (``the universe") consisting of a black hole of mass~$M$ and a bucket of water
(which carries energy and entropy) 
outside the black hole horizon. Now imagine a process where one throws the water into the black hole, say in a 
perfectly radial direction so as not to generate angular momentum. 
After the water has entered the horizon, the external region loses some 
energy~$E_\text{water}$ and some entropy~$S_\text{water}$. 
Once the perturbations of the system have died down, the system settles into a new state
with a new black hole solution and nothing outside. 
The mass of the black hole can be measured from the outside and it increases to~($\wt M = M+E_\text{water}$), 
such that the total energy of the system remains the same (as it should by the principle of conservation of energy!). 
On the other hand, it would seem that the total entropy of the system decreases by~$S_\text{water}$---thus 
violating the second law of thermodynamics---unless one assigns an intrinsic entropy to the black hole 
itself. 

Now, if a black hole has entropy, and given that it has energy (= its mass~$M$), it must have a temperature, 
according to the first law of thermodynamics.
Building on the fact that the black hole area always increases in physical processes~\cite{Bardeen:1973gs}, 
Bekenstein further argued that the thermodynamic entropy of black holes must be proportional to 
the area of its event horizon. 
In a tour de force, Hawking then set up a thought experiment involving scattering of  
particles (or, equivalently, waves in the quantum theory) off a black hole using the formalism of quantum field theory. 
Within this setting he calculated the ratio of the amplitude of the reflected wave and the incident wave, which directly
leads to the rate of radiation---from which he could deduce the temperature using the Planck 
distribution formula, and therefore a precise formula for the entropy. 

These considerations result in the following succinct formula  
for the thermodynamic entropy of a BH, 
that applies universally to any BH solution in general relativity\footnote{In fact this entropy formula 
has a remarkably wide range of applicability, and holds in theories of 
general relativity coupled to matter fields in various dimensions (with the appropriate
generalization of the horizon area).}: 
\be \label{BHentropy}
S^{\!\text{class}}_{\!\text{BH}} \= k_B \, \frac14  \frac{A_H}{\ell_\text{Pl}^2} \,, 
\qquad \ell_\text{Pl}^2 \= \frac{G \hbar}{c^3}\,,
\ee
where~$A_H$ is the area of the BH horizon.\footnote{The superscript refers to the fact that this is a 
semi-classical formula, which will later be promoted to a quantum formula.} 
The Bekenstein-Hawking black hole entropy formula~\eqref{BHentropy} is one of the most profound 
equations in theoretical physics, involving the three fundamental constants: the speed of light~$c$, 
Planck's constant~$\hbar$, and Newton's gravitational constant~$G$, which determine the fundamental 
scales of special relativity, quantum mechanics, and gravitation, respectively, as well as the Boltzmann 
constant~$k_B$ which determines the scale of thermodynamic entropy.
These constants will often be suppressed in the following formulas for ease of presentation.

From the theory of quantum-statistical mechanics, one knows that the entropy of a physical system is 
really a statistical property arising from an underlying collection of quantum-mechanical microscopic 
states (or \emph{microstates}). 
This property is quantified by the Boltzmann equation which expresses the thermodynamic 
entropy as the logarithm of the number of microscopic states in which the system can exist for a 
given macroscopic state. Now, upon combining the black hole entropy formula with the Boltzmann 
equation one obtains the Boltzmann equation for black holes
\be \label{BHBoltzmann}
k_B \log d_\text{micro} \= S^{\!\text{class}}_{\!\text{BH}} + \dots \,.
\ee
The dots here indicate that the Boltzmann equation is an approximate equation, which is 
valid in the so-called \emph{thermodynamic limit} of very large black hole size. 

One thus reaches the profound conclusion that a black hole
should be made up of a collection of a large number~$d_\text{micro}$ of microscopic states.

\vskip 0.2cm

Finding a theory of quantum gravity---a theory which brings together Quantum Mechanics and General Relativity 
into one consistent framework---has remained an important open question of fundamental physics for the last 60 years.
One of the reasons it is such a hard problem is the lack of an experimental guide due to constraints on 
current technology. 
In this situation it is very useful to have a quantitative criterion which can be used to test or falsify a given theory.
The Bekenstein-Hawking entropy formula plays this role: 
a consistent theory of quantum gravity should be able to produce a microscopic 
quantum-statistical ensemble of black hole microstates which satisfies
the Boltzmann equation~\eqref{BHBoltzmann} for black holes.

\section{Two pictures of black holes in string theory \label{sec:BHstringtheory}}

\begin{figure}\centering
\includegraphics[width=14cm]{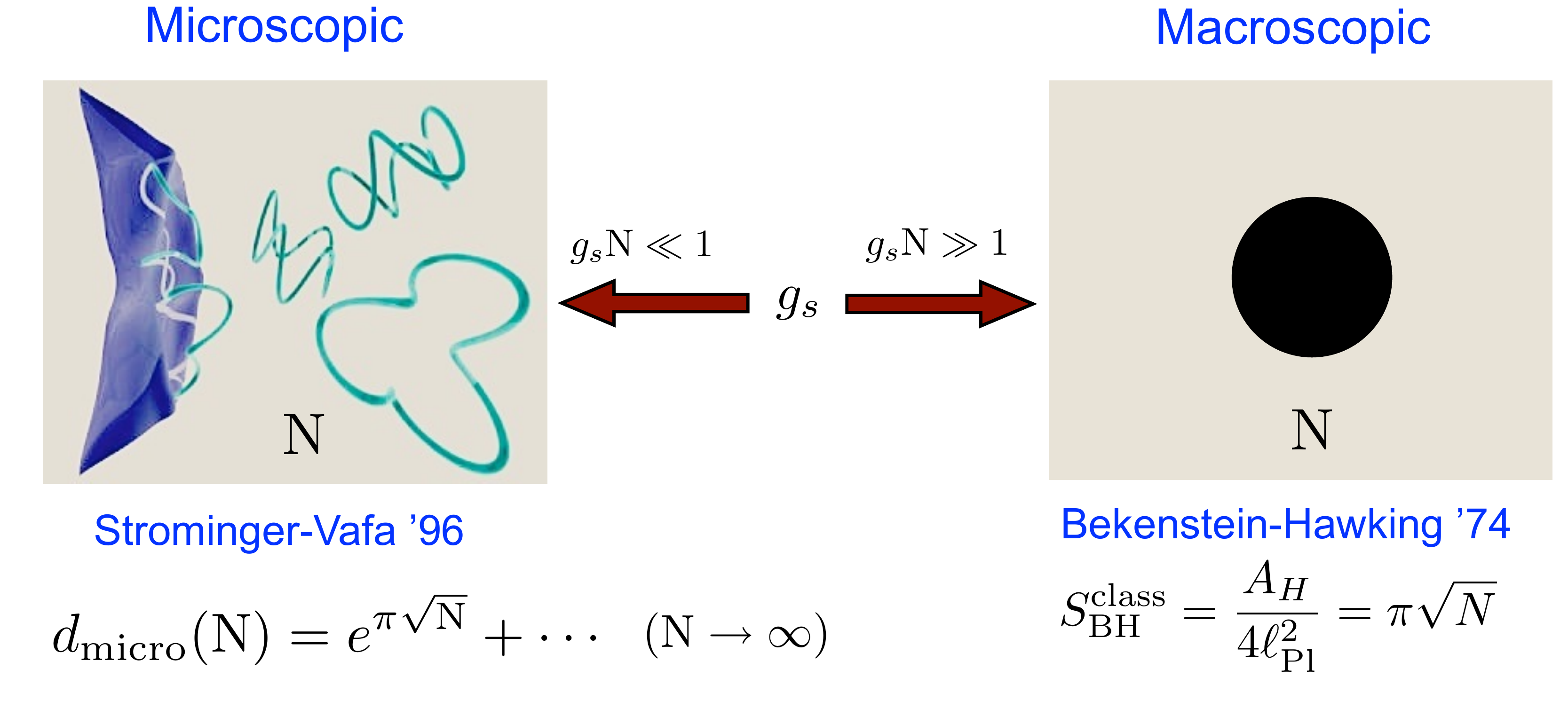}
  \caption{Microscopic and macroscopic pictures of a black hole in string theory}
  \label{BHmicmac}
\end{figure}

A series of developments in the 1990s pioneered by~\cite{Sen:1995in} and~\cite{Strominger:1996sh},
building on previous work by~\cite{tHooft:1990fkf,Susskind:1993ws,Susskind:1994sm},
led to a quantum-statistical explanation of the thermodynamic entropy of black holes in string theory. 
The basic idea is that there are actually two pictures of a black hole in string theory
(see Figure~\ref{BHmicmac}), which allows us to separately calculate and compare 
its thermodynamic and statistical entropy.

Let us first consider the basic physical question:  
what determines whether an object of mass~$M$ is a black hole or not?  
The criterion, roughly speaking, is whether the intrinsic size of the object is smaller or larger than 
the size of a black hole of the same mass. 
Consider, for the sake of simplicity, spherically symmetric objects. Then 
if the radius of the object is larger than the \emph{Schwarzschild radius}~$R_\text{Sch}=2 GM$, 
then it is \emph{not} a black hole, while if its radius is smaller than the Schwarzschild radius then 
there is pure vacuum outside the horizon and one cannot distinguish the object from a black hole. 
(Recall that one cannot observe anything behind the horizon.)
For composite objects like stars, the intrinsic size is determined by the radiation pressure of 
electrons, and the above logic leads to the so-called Chandrashekar 
limit.\footnote{See~\url{https://en.wikipedia.org/wiki/Chandrasekhar_limit}.}

Now we apply this criterion to string theory. 
The consistency of supersymmetric string theory imposes that space-time is 
ten dimensional. In order to obtain an effective theory in~$\IR^{1,3}$, six of the dimensions 
should describe a compact manifold. The resulting geometry is called a string compactification,
and it determines the spectrum of 
particles and their interactions at low energies. 
(See e.g.~the classic textbook~\cite{Green:1987sp} for more details.)
The spectrum is characterized by the conserved charges of the theory, also called~\emph{quantum numbers}, 
which are quantized as integer multiples of fundamental units of charge in the quantum theory. 
In the simplest case the spectrum is labelled by a single quantum number~$N \in \IN$.

The gravitational coupling (the effective Newton's constant as appears in all equations) in string theory 
is determined by the value of the string coupling~$g_\text{s}$ as~$G=g_\text{s}^2 \ell_\text{s}^2$ 
where~$\ell_\text{s}$ is the fundamental scale of a string. 
In this context, the criterion of whether the collection of states labelled by~$N$ is a black hole or not 
takes the following form. 
When~$g_\text{s}N \gg 1$ the effective description is that of weakly-coupled general relativity interacting with  
a specified set of matter fields, and one has  
a black hole solution of this effective theory with quantum number~$N$. This is the \emph{macroscopic} picture. 
When~$g_\text{s}N \ll 1$, there is a completely different dual microscopic picture: 
one has a set of fundamental excitations of string theory with the same quantum number~$N$, 
consisting of bound states of fluctuations of strings, 
branes, and other fundamental objects of string theory.

The microscopic picture gives us the possibility of calculating the statistical entropy. 
However, for a given value of~$g_s$, only one of the pictures should hold,
and one still cannot make a comparison between the two types of entropies.\footnote{Note, however, 
that it leads to the idea of a transition between
black holes and microstates~\cite{Horowitz:1996nw}.} 
The breakthrough of~\cite{Sen:1995in,Strominger:1996sh} was to consider the black hole entropy 
problem in the context of \emph{supersymmetric compactifications} of string theory.
In a class of such compactifications, the string coupling constant~$g_s$ turns out to be not fixed by the 
equations of motion, and therefore is a tunable parameter.  

Now one can test whether black hole entropy is a statistical entropy as follows. 
Start with a particular supersymmetric compactification of string theory which admits black hole 
solutions with some set of conserved charges.
Tune the coupling so that the effect of gravity is arbitrarily small and, therefore, the fluctuations of 
strings and branes can be described using usual quantum field theoretic methods. 
Count the number of microstates~$d_\text{micro}$ in the theory as a function of 
the quantum numbers. 
Then crank up the coupling and therefore~$G$, so that the horizon radius is much larger than
the fundamental scale of the microscopic constituents. 
Now the collection of microstates gravitate and form a black hole. 
Measure its horizon area and obtain the thermodynamic entropy~$S^{\!\text{class}}_{\!\text{BH}} $.
Finally, compare~$S^{\!\text{class}}_{\!\text{BH}} $ and~$\log d_\text{micro}$ as a function 
of the quantum numbers.

However, one is faced with yet another problem---the number of states of the theory can change 
as one changes the coupling, so it is not clear that one should be comparing the entropy of states calculated 
at weak coupling with the black hole entropy calculated at strong coupling. 
The resolution of~\cite{Sen:1995in, Strominger:1996sh}  is to consider a 
special set of states which exist in supersymmetric theories called BPS (after~\cite{Bogomolny:1975de, Prasad:1975kr}) 
states which are ``protected", which is to say that the number of BPS 
states---counted with a certain weighting, called the supersymmetric index---does not change under 
change of small parameters of the theory and their entropy can be compared with the entropy of supersymmetric BHs. 
These ideas are explained in the following section.

\section{Supersymmetric states and supersymmetric black holes \label{sec:susystatesBH}}

In order to explain the supersymmetric index it is useful to first briefly review the 
basic ideas underlying supersymmetry. 
All particles in nature comes in two types---bosons and fermions. 
The bosonic or fermionic nature of particles determines their fundamental properties such as their 
spin---which can be integer multiples of~$\hbar$ for bosons and integer+$\half$ multiples for fermions, 
and statistical properties: bosons tend to cluster in the same state, i.e.~the probability for a boson to be in a 
certain quantum state increases with the number of bosons already present in that state, while fermions repel 
each other, i.e.~the probability of two fermions to be in the same quantum state is zero. 
Supersymmetry is a symmetry, first proposed by particle physicists in the 1970s, that relates fermions 
and bosons.\footnote{Although it has not been observed in nature, 
it is very useful to construct model systems where one can perform controllable analytic calculations 
(which are usually out of reach) in order to probe physical properties of quantum theories e.g.~\cite{Seiberg:1994rs}. 
Here we use it to test the Boltzmann equation for black holes.}

The main ideas of supersymmetric state counting can be illustrated in the context of a 
simple type of system called supersymmetric quantum mechanics~\cite{Witten:1982df}.
Recall that a quantum mechanical system has a Hilbert space~$\CH$ whose elements are called (quantum) states, 
and a self-adjoint operator~$H$ on the Hilbert space, called the Hamiltonian. 
The eigenvalues of the Hamiltonian are interpreted as possible energies of the states. 
Now, consider a quantum mechanical system whose Hilbert space~$\CH$ consists of states 
which can be bosonic or fermionic. The \emph{fermion number operator} denoted~$(-1)^F$ 
is a multiplicative operator defined to have value~$+1$ for bosonic states and~$-1$ for fermionic states. 
The supersymmetric quantum mechanical system comes equipped with a pair of complex conjugate 
fermionic operators~$Q, \overline{Q}$, called the supercharges, 
which are nilpotent (i.e.~$Q^2 = \overline{Q}^2 = 0$), and obey the algebra
\be \label{susyalg}
\{ Q, \overline{Q}\} \= H \,, \qquad [H,Q] \= 0 \,, \qquad [H,\overline{Q}] \= 0 \,.
\ee
The spectrum of states of the supersymmetric system falls into representations of this algebra, which come in two types.
The first type is a two-dimensional \emph{long representation}~$(| b\rangle, | f \rangle)$ with~$Q | b \rangle = | f \rangle$,
with~$Q | f \rangle = 0$ by the nilpotency of~$Q$.
Since~$Q$ has fermion number~$-1$, the two states have opposite fermion numbers, which are denoted by   
 $b$ (bosonic) and  $f$ (fermionic).
If~$ | b \rangle$ is an eigenstate of~$H$ with eigenvalue~$E$, the algebra~\eqref{susyalg} 
implies that~$E \ge 0$ and that~$ | f \rangle$ is also an eigenstate with eigenvalue~$E$.
The second type is a one-dimensional  \emph{short representation} $|  \cdot \rangle$, which can be either 
bosonic or fermionic, which obeys~$Q | \cdot \rangle = \overline{Q} | \cdot \rangle = 0$. 
The the algebra~\eqref{susyalg} implies that~$| \cdot \rangle$ is an eigenstate of~$H$ with eigenvalue~$E=0$. 
Thus the complete spectrum of the theory consists of positive energy states which come in boson-fermion pairs,
and zero energy states which need not be paired. 

The basic supersymmetric index, called the \emph{Witten index}~\cite{Witten:1982df},  is defined as
\be
Z \= \Tr_{\CH} \, (-1)^F \, \rme^{- \beta H} \,.
\ee
This quantity is similar to the thermal partition function that one encounters in statistical physics, with~$\beta$ being 
the inverse temperature. In particular, states with large energies are suppressed by the exponential factor, which  
usually leads to a convergent trace. 
There is, however, a big difference compared to the thermal partition function. Since states with~$E>0$
come in pairs with opposite values of~$(-1)^F$, they do not contribute to the spectrum.
The only contribution to the index comes from zero-energy states, i.e.
\be \label{Windex}
Z \= n_0^b - n_0^f \,,
\ee
the difference between the number of bosons and fermions with zero energy, which, in particular, is
independent of~$\b$! 

Now consider the effect of changing some parameter of the system in a supersymmetric manner, 
that is to say, the supercharges and the Hamiltonian are functions of the parameter, but obey 
the algebra~\eqref{susyalg} for all values of the parameter. 
The energy levels (the eigenvalues of the Hamiltonian) 
correspondingly change, but bosons and fermions with~$E>0$ are still paired. 
Any change in the~$E=0$ spectrum consists of states moving down from~$E>0$ to~$E=0$,
or leaving it in the opposite direction, but this can only happen in boson-fermion pairs. 
Thus, although the total number of (bosonic + fermionic) states at~$E=0$ could change, 
their difference is invariant.

In this simple example, one sees that the supersymmetric index only receives contributions from 
the one-dimensional zero-energy representations, which are annihilated by the supercharge operators. 
In the more general setting of supersymmetric quantum field theories, one has multiple supercharges.
The types of representations of the supersymmetry algebra in this case are labelled by the 
fraction of the total number of supercharge operators that are annihilated in that representation. 
This fraction can range from~0, the so-called non-BPS or long representations, 
to~1, which are the zero energy vacuum states as in the simple example above. However, now one 
also has BPS states with non-zero energy, which are annihilated by only some of the 
supercharges. Correspondingly, one has more general types of supersymmetric indices, 
which receive contributions only from states that are annihilated by at least the same fraction of the supercharges.
The precise details of these supersymmetric indices depends on the full symmetry algebra of the theory.
In the context of asymptotically flat space (as discussed in this article) the symmetry algebra is super-Poincar\'e, 
in which case the relevant indices are called helicity supertraces~\cite{Kiritsis:1997gu}.

\medskip

\ndt {\bf The micro/macro picture revisited}

Now consider superstring theory and a particular supersymmetric index that counts a certain type of BPS state. 
Since the index does not change under changes of coupling, one can revisit the idea of Figure~\ref{BHmicmac} 
and ask what happens to these states in the macroscopic regime~$g_\text{s}N \gg 1$? The complete 
answer depends on the details of the system but, happily, there are situations where 
one has a black hole solution of string theory with the same charges as those of the original microscopic quantum 
states.
In such a situation, one can try to compare the microscopic and macroscopic pictures more carefully. 

Upon reviewing the above logic carefully, one notices that the original problem seems to have been changed at two levels. 
Firstly, the states that are being considered are annihilated by a certain fraction of supercharges, and this 
should be reflected in the gravitational theory. Indeed, the effective macroscopic theory---called supergravity---also 
has a notion of supersymmetry that applies to the gravitational variables, and there are black hole solutions 
to supergravity---called BPS black holes---that are annihilated by some fraction of the set of supercharges.  
BPS black holes are a very special type of black hole solutions. 
One of their peculiarities is that they are necessarily have zero temperature. 
In this respect they are different from Schwarzschild black holes (as initially mentioned in the introduction). 
Nevertheless, BPS black holes have an event horizon and do carry entropy!\footnote{The apparent violation of the 
lore/notion of the third law of thermodynamics is avoided because the supersymmetry of such solutions
allows for a large degeneracy.} 

The second level at which the initial problem seems to have changed is that one is no longer 
discussing the total number of states (and the consequent Boltzmann entropy), but a modified counting of 
the index of the type~\eqref{Windex}. 
In the early days of black hole microstate counting in string theory this aspect 
was not quite understood fully and it was even questioned whether the agreement obtained by~\cite{Strominger:1996sh} 
was only some kind of accidental or approximate agreement in the limit of large charges. 
More recent developments have made it clear that this is not an accident and 
there is a more detailed and beautiful mechanism underlying the system. 

The basic physics picture is as follows. 
Typically, boson-fermion pairs do pair up and leave the spectrum upon reaching strong coupling. 
Assuming for the moment that the initial microscopic system has more bosons than 
fermions, one is left with a final ensemble 
in which~$n^f=0$ and the index only counts bosonic states, i.e.~a genuine count in the sense of Boltzmann. 
On the macroscopic side, it is a non-trivial fact~\cite{Sen:2009vz}, \cite{Dabholkar:2010rm}, \cite{Iliesiu:2022kny}
that the BPS black holes are indeed made up of 
only bosonic states (this is sometimes referred to as a vanishing theorem)!\footnote{
The bosonic nature of the states has been verified in~\cite{Dabholkar:2010rm} for various examples of 
BPS black holes in string theory. The theoretical argument assumes the decoupling of the quantum states 
of the BH from the fluctuations of the fields outside the horizon. This was recently shown to hold for supersymmetric 
BHs in string theory~\cite{Iliesiu:2022kny}.}  
Thus, one has a logic of the following sort: 
\begin{center}
microscopic index  =  gravitational index  =  BH entropy , \\
\qquad \qquad (supersymmetry)   \qquad (vanishing theorem) \;
\end{center}
so that the supersymmetric index 
should indeed capture the Boltzmann entropy of supersymmetric black holes. 

\medskip

With the formalism in place, one proceeds to calculate the supersymmetric index at large charges and check 
whether it agrees with the Bekenstein-Hawking area formula for the entropy.

\section{The counting of black hole microstates in string theory \label{sec:BHmicrostates}}

As mentioned in the introduction, black hole microstates in string theory
are quantum-mechanical bound states of fluctuations of fundamental objects of string theory like strings and branes. 
We first consider a toy model, consisting of a string wrapping a circle of radius~$R$, 
which nicely illustrates a number of features of the microscopic counting of black hole 
microstates in string theory. 

The string is a~$1$-space-dimensional object evolving in time, whose points are 
labelled by the coordinates~$x^0$~(time) and $x^1$~(space). 
The position on the circle is parameterized by the map~$X(x^0,x^1) \in \IR/2 \pi \ii R \IZ$.  
 The classical equation of motion on the free string is the free Laplacian equation 
(with~$\p_i \equiv \p/\p x^i$), 
\be
\bigl(\p_0^2 - \p_1^2 \bigr) X(x^0,x^1) \= 0 \,.
\ee
The general solution of this equation is
\be
X(x^0,x^1)  \= \frac{n}{R} x^0 + w R  x^1   + \, \sum_{k \in \IZ \atop k \ne 0} \,  \frac{\ii}{\sqrt{k}}
 \Bigl(  \a_k \, \rme^{-\ii k (x^0 - x^1)}  \,+\,   \wt \a_{k} \, \rme^{-\ii k (x^0 + x^1)} \, \Bigr) \,.
\ee
Here~$n,w \in \IZ$ are interpreted as the momentum of the center of mass of the string
around the circle and the winding of the string around the circle, respectively. The terms 
proportional to~$\rme^{-\ii k (x^0 - x^1)}$ and~$\rme^{-\ii k (x^0 + x^1)}$ are called left- and right-moving
excitations, respectively. We focus on the left-moving excitations, and the right-movers are treated in an analogous manner.
In the classical-mechanical theory, the coefficients~$\a_k$ are complex numbers
obeying~$\overline{\a_k}=\a_{-k}$. 
In the quantum theory, the field~$X$ and therefore the coefficients~$\a_k$, $\wt \a_k$ are promoted 
to operators on a Hilbert space obeying~$\a_k^\dagger=\a_{-k}$. 
The quantization process leads to the following commutation relations
\be \label{RLops}
[\a_{k}, \a_{-k}] \=  1  \,, \quad  k \= 1, 2, 3, \dots \,,
\ee
with all other commutators vanishing. 

The algebra obeyed by the pair~$(\a_k, \a_{-k})$, defines the quantum bosonic oscillator, 
which is one of the simplest known quantum-mechanical systems.
The Hilbert space~$\CH_k$ associated to this system is described as follows. 
There is a special \emph{vacuum} state~$|0\rangle_k$ 
of this system obeying~$\a_{k}|0\rangle_k = 0$, $k>0$, and the Hilbert space is given by the span over~$\IC$ 
of the following tower of states,
\be \label{statesmk}
|m \rangle_k \= \a_{-k}^{m}|0\rangle_k \,, \qquad m = 0, 1,2, \dots \,.
\ee
The Hamiltonian of this system is given by
\be
H_k \= k \bigl(\a_k \, \a_{-k}+\tfrac12 \bigr) \,, 
\ee
and the energy eigenvalues of the states~\eqref{statesmk} can be easily calculated 
using the commutation relation in~\eqref{RLops} to be
\be
H_k |m \rangle_k \= k(m+\tfrac12) |m \rangle_k \,.
\ee
The thermal partition function of this harmonic oscillator is given by
\be \label{Hkpartfn}
\Tr_{\CH_k} \, \rme^{-\b H_k} \= 
\sum_{m=0}^\infty \rme^{-\b k (m + \frac12)}
\= \frac{\rme^{-\b k/2} }{1-\rme^{-\b k}} \,.
\ee

The full left-moving Hilbert space~$\CH$ of the theory is the Fock space 
built out of these individual Hilbert spaces~$k \in \IZ^+$, 
and the total left-moving Hamiltonian is given by~$H = \sum_{k=1}^\infty H_k $ 
where~$H_k$ is now interpreted to act as the identity operator on all the Hilbert subspaces~$k' \neq k$. 
The thermal partition function on this Hilbert space is the product over all~$k>0$ of 
the thermal partition functions~\eqref{Hkpartfn}. 
The exponents of the numerators of~\eqref{Hkpartfn} add up to give the energy eigenvalue of the vacuum state,
which, according to the above discussion, is naively given by~$\sum_{k=1}^\infty \frac{k}2$.
This sum is interpreted using the zeta-function regulator\footnote{Recall that~$\zeta(s) = \sum_{k=1}^\infty k^{-s}$
is defined for~$\Re(s) > 1$, and can be analytically continued to a meromorphic function on the whole complex plane.}  
to be~$\zeta(-1) = -1/24$. 
Setting~$\b = - 2 \pi \ii \tau$,~$q=\rme^{2 \pi \ii \tau}$, one obtains, 
\be \label{FreebosPfn}
\begin{split}
 \Tr_{\CH} \, q^{H} & \= q^{-\frac{1}{24}} \prod_{k=1}^\infty \, \frac{1}{1-q^k}
 \= \sum_{n=0}^\infty \, p(n) \, q^{n-\frac{1}{24}} \\
& \= q^{-\frac{1}{24}} \bigl( 1 + q + 2 q^2 + 3 q^3 + 5 q^4 + \dots \bigr) \,. 
\end{split}
\ee
The coefficient~$p(n)$ in the above expansion is the number of partitions of the integer~$n$, 
and can be expressed as the integral 
\be \label{pncontint}
p(n) \= \frac{1}{2 \pi \ii} \oint_C \frac{dq}{q^{n+1}} \,  \prod_{k=1}^\infty \, \frac{1}{1-q^k}  
\ee
over a contour lying inside the unit disk and surrounding the origin.  
The asymptotic behavior of~$p(n)$ as~$n \to \infty$ is well-known, and 
can be calculated using the well-known relation of the above infinite
product to the Dedekind~$\eta$-function 
\be \label{Dedekind}
\frac{1}{\eta(\tau)} \= q^{-\frac{1}{24}} \prod_{k=1}^\infty \, \frac{1}{1-q^k} \,, \qquad q=\rme^{2 \pi \ii \tau} \,.
\ee
As~$n \to \infty$, the integral~\eqref{pncontint} is dominated by the singular behavior of the integrand 
near the unit circle, which is governed by the amazing symmetry property obeyed by the~$\eta$-function,
which is expressed as the functional equation 
\be
\eta(-1/\t) \= \sqrt{-\ii\t} \, \eta(\t) \,.
\ee
Using this, one obtains 
\be
\log p(n) \sim 2 \pi \sqrt{n/6} + \dots 
\ee

\bigskip

The actual counting of supersymmetric microstates in string theory is more complicated 
than the above toy example, but retains many of its basic features. 
As mentioned above, 
we consider compactifications of ten-dimensional string theory on a six-manifold~$\CM_6$
to obtain a geometry of the form~$\IR^{1,3} \times \CM_6$. 
There are, in fact, multiple versions of string theory, and we focus 
on one theory called Type IIB string theory. 
It is known that Type IIB string theory compactified on 
a Calabi-Yau 3-fold leads to a supersymmetric macroscopic theory in four dimensions described by 
General Relativity coupled to multiple gauge fields, scalar fields, and fermions. 

This theory admits supersymmetric black hole solutions, labelled by their charges under all the gauge 
fields of the theory. 
The corresponding microscopic states involves excitations of 
string theory called branes, which are higher-dimensional generalizations of strings,
wrapping cycles of the Calabi-Yau manifold.
Counting the number of such microscopic states for a given vector of charges 
is a complicated problem, and in general one only has estimates for this number given in~\cite{Maldacena:1997de}. 
However, it can be solved exactly when the Calabi-Yau is sufficiently simple. Typically, in these cases, 
the generating functions of microstates consist of combinations of Dedekind~$\eta$-functions and 
classical~$\theta$-functions. 

The simplest situation is when the Calabi-Yau manifold is~$T^6$. The resulting macroscopic theory is called~$\CN=8$
supergravity, which admits supersymmetric black hole solutions labelled by one positive integer~$N$ (cf.~Fig.~1), 
whose Bekenstein-Hawking entropy is given by
\be \label{N8BHent}
S^\text{class}_\text{BH}(N) \= \pi \sqrt{N} \,,
\ee
The corresponding microstates are described by left-moving oscillations of an effective string wrapping some cycles in~$T^6$.
The single circle in the toy example is now replaced by multiple free bosons and also fermions, 
and the generating function of the microscopic number of states\footnote{As explained above, one really calculates
the microscopic index. In this case, it is known that the microscopic index is related to the number of black hole 
microstates as~$(-1)^{N+1} d_\text{micro}(N)$, due to fermionic zero-energy modes that one has to factor 
out~\cite{Sen:2009vz},~\cite{Dabholkar:2010rm}. 
The resulting generating functions~$(-1)^a \theta_a(\tau)/\eta(\tau)^6$, $a=1,2$ transform as a vector under~$SL(2,\IZ)$ 
modular transformations, and they can be assembled into a single Jacobi form~\cite{EichZag}.}
is given by
\be \label{dmicFexp}
\sum_{N= -1 \atop N \; \equiv \; (a +1) \; \text{mod} 2}^\infty  \, d_\text{micro}(N) \, q^{N/4} \=  \frac{\theta_a(\t)}{\eta(\tau)^{6}}\,, 
\ee
where~$\theta_a(\tau)=\sum_{n \in \IZ+a/2} q^{n^2}$, $a=0,1$. 
The asymptotic behaviour of states can be calculated in a similar manner as for integer partitions in~\eqref{pncontint},
using the symmetry property of the~$\theta$ and~$\eta$ functions relating the behaviors at~$\t$ and~$-1/\t$,
and the result is, as~$N \to \infty$, 
\be \label{N8micro}
\log d_\text{micro}(N) \; \sim \; \pi \sqrt{N} + \dots \,.
\ee

The agreement of the statistical (Boltzmann) entropy of microstates~\eqref{N8micro} and the thermodynamic entropy 
of black holes~\eqref{N8BHent}, as pictured in Figure~1, is a major success of string theory. 
The generating function of the microscopic degeneracy of states in this example is built out of 
very special functions like the Dedekind~$\eta$-function and the~$\theta$-functions,
whose symmetry under~$\t \to -1/\t$ is very useful to estimate the growth of the degeneracy of states. 
These functions are examples of a very special type of functions called modular forms which 
are symmetric under an infinite group of symmetries called the modular group.

\section{Modular forms and the Hardy-Ramanujan-Rademacher expansion \label{sec:BHMF}}

In this section we switch gears and discuss some mathematical aspects of the theory 
of modular forms. In the following section we discuss some consequences of this theory which have implications on 
the quantum aspects of black hole entropy. 

\medskip

The theory of modular and automorphic forms is a vast field of mathematics, and we only discuss a very 
small part of it which is needed for the sequel. In particular, we discuss holomorphic functions of 
one variable~$\t \in \IH$, the upper half-plane. Recall that the modular group~$SL(2,\IZ)$, 
the group of $2\times 2$ matrices with integer entries and determinant one, acts on~$\IH$ by fractional linear 
transformations.
The central defining property of a modular form~$f(\tau)$ is its transformation under this action, 
\be \label{modtrans}
f \biggl( \frac{a\t+b}{c\t+d} \biggr) \= (c\t+d)^k \, f(\t) \,,  \qquad 
 \biggl( \, \begin{matrix} a & b \\ c & d \end{matrix} \, \biggr) \in SL(2,\IZ) \,.
\ee
Here~$k$ is called the~\emph{weight} of the modular form.
In writing~\eqref{modtrans}, the weight is assumed to be integer.  
The theory of modular forms exists for half-integer and even more general values of~$k$, and 
also for congruence subgroups of~$SL(2,\IZ)$. 
The details in each case vary: for example, there could be roots of unity appearing in the transformation law.

Taking the~$SL(2,\IZ)$ matrix in~\eqref{modtrans} to be~$\left( \begin{smallmatrix} 1 & 1 \\ 0 & 1 \end{smallmatrix} \right)$,
one obtains the periodicity property~$f(\t+1) = f(\t)$ which, together with the holomorphy, leads to 
 the Fourier expansion 
\be \label{MFFourier}
f(\t) \= \sum_{n \in \IZ} d(n) \, q^n \,, \qquad q \= \rme^{2\pi \ii \t} \,.
\ee 
This expansion, called the~$q$-expansion in this context, is at the heart of many of the 
relations of modular forms with other problems in mathematics and physics. In particular, the 
coefficients~$d(n)$ are often integers, and can sometimes be interpreted as (virtual) dimensions 
of certain Hilbert spaces. Indeed, in the context of our main interest here, the coefficients are supersymmetric 
indices corresponding to black hole microstates in string theory, as explained in the previous section.

When the sum in~\eqref{MFFourier} is restricted to~$n \ge 0$, the corresponding functions are   
holomorphic on the upper half-plane, and are called holomorphic modular forms. 
Those with the stronger restriction~$n >0$ are called cusp forms. 
However, it can be proved that the growth of  
coefficients~$d(n)$ as~$n \to \infty$ of holomorphic modular forms is bounded by 
a polynomial function. 
The exponential growth of~$d(n)$ that is characteristic of black hole microstates appears in modular forms 
with negative powers of~$n$ in its~$q$-expansion (such functions are called \emph{weakly holomorphic}) 
in which case the function itself diverges as~$\t \to \ii \infty$.  
In order to obtain a controllable analytic theory of such functions, 
one restricts the sum in~\eqref{MFFourier} in the negative direction to~$n \ge n_0$ for some fixed~$n_0<0$. 
This is indeed the case for the examples in the previous section, 
e.g.~$n_0=-1/24$ in~\eqref{Dedekind}\footnote{In fact, it is $1/\eta(\tau)^{24}$ that is a modular form on~$SL(2,\IZ)$, 
and the transformation~\eqref{modtrans} of~$1/\eta(\tau)$ involves a $24^\text{th}$ root of unity.}.

\bigskip

\ndt {\bf Asymptotic expansion of coefficients of modular forms}

The leading asymptotic behavior of the coefficients of weakly holomorphic modular forms can be estimated quite easily,
as seen in the examples of the previous section. 
In fact, the modular symmetry is much more powerful,
and leads to an exact analytic formula for the coefficients of functions of the type~\eqref{Dedekind} 
and~\eqref{dmicFexp}, called the \emph{Hardy-Ramanujan-Rademacher formula} after 
\cite{HardyRam},~\cite{Rademacher}. The formula for the coefficients of~\eqref{dmicFexp} is: 
\be\label{RadexpC} 
 d_\text{micro}(N) \=   2{\pi} \, \Bigl( \frac{\pi}{2} \Bigr)^{7/2} \, \sum_{c=1}^\infty 
  c^{-9/2} \, K_{c}(N) \; \wt I_{7/2} \Bigl(\frac{\pi \sqrt{N}}{c} \Bigr)  \,.
\ee
Here~$\wt{I}_{\rho}$, a modification of the standard $I$-Bessel function,
is given by the following integral formula, 
\begin{equation}\label{Besselintrep}
 \wt{I}_{\rho}(z)\=  \Bigl(\frac{z}{2} \Bigr)^{-\rho} I_{\rho}(z) 
 \=\frac{1}{2\pi \ii}\int_{\epsilon-\ii\infty}^{\epsilon+\ii\infty} \, 
 \frac{d\sigma}{\sigma^{\rho +1}}\exp \Bigl(\sigma+\frac{z^2}{4\sigma} \Bigr) \, .
\end{equation}
The Kloosterman sum~$K_c$  in~\eqref{RadexpC} is a sum over phases, which is a class of functions 
appearing in many number-theoretic contexts. The reader is referred to~\cite{Iliesiu:2022kny} 
for the details of this example.

\section{Quantum corrections to black hole entropy \label{sec:QuanCorr}}

The discussion of Sections~\ref{sec:BHmicrostates} and~\ref{sec:BHMF} showed 
that the macroscopic, thermodynamic black hole entropy~\eqref{BHentropy} is indeed 
reproduced by the statistical Boltzmann entropy of the corresponding black hole microstates
in the thermodynamic limit when the black hole size becomes very large, i.e.,
\be \label{micmacagree}
\log d_\text{micro}(N) \= S^\text{class}_\text{BH}(N) + \dots \,,
\ee
where the integer~$N$ is the quantized charge of the black hole. 
This match is one of the big successes of string theory.

The quantities appearing on the two sides of~\eqref{micmacagree} have very different 
mathematical natures as well as physical origins: 
the left-hand side is defined by the arithmetic function~\hbox{$d_\text{micro}: \IN \to \IN$} 
that counts the dimension of the microscopic Hilbert space, and has its origin in quantum-statistical mechanics,  
while the right-hand side is defined by the smooth real function~\hbox{$A_H: \IR^+ \to \IR^+$} that measures 
the area of the BH horizon in semi-classical General Relativity, and is intrinsically geometric in origin.  
It is therefore very surprising that the two quantities are equal to each other! 
Is there a deeper explanation of this equality? 

One can try to sharpen this question as follows. The dots on the right-hand side of~\eqref{micmacagree} 
refer to the fact that the area formula is really an approximation, as~$N \to \infty$, to the logarithm of 
the exact integer given by the left-hand side.\footnote{The difference in the nature of the domain of the 
two functions is well-understood: it is explained by the fact that  
the charge of the black hole in classical general relativity is continuous, while in the 
quantum theory it should be quantized in units of the fundamental charge. 
}  
One is thus led to ask: is it an approximation to an exact formula? 
Note that the left-hand side is perfectly well-defined for any integer~$N$, so the question 
is whether there is an exact quantity defined in the macroscopic, gravitational  
theory, which agrees with the integer~$d_\text{micro}(N)$ appearing on the left-hand side.
The pursuit of the answer to this question has led to a very beautiful story spanning 
a few decades of research in string theory, and has resulted in non-trivial relations with number theory
referred to in the introduction. 
The rest of this section presents a sketch of the answer using the prototype system mentioned
in the previous section, along with references to the literature for more detailed explanations. 

The Hardy-Ramanujan-Rademacher expansion~\eqref{RadexpC} already
suggests a possible answer to the above questions at a mathematical level: the integer~$d_\text{micro}(N) $ 
is expressed as the sum of a series of analytic functions evaluated at discrete values of the argument
controlled by~$N$. This series can be interpreted precisely as arising from quantum 
gravitational corrections to the Bekenstein-Hawking area formula for the thermodynamic entropy of the black hole,
as anticipated in~\cite{Dijkgraaf:2000fq, Ooguri:2004zv, Dabholkar:2005by} and as explained below.

Recalling that~$I_\rho(z) \sim \rme^z$, $z \to +\infty$, it is clear from the form of the expansion~\eqref{RadexpC} 
that the~$c>1$ terms are all exponentially suppressed compared to the~$c=1$ term. 
One already saw that the leading asymptotic growth of~$d_\text{micro}(N)$, given in Equation~\eqref{N8micro},
agrees precisely with the Bekenstein-Hawking entropy formula~\eqref{N8BHent} for the BH entropy. 

The starting point for the quantum corrections to this formula 
is the observation that the spacetime geometry has a special form
near the horizon of supersymmetric black holes. 
In particular, the manifold is a product of~AdS$_2$ (two-dimensional constant negative curvature spacetime)
and a compact manifold (S$^2$ for the BH under consideration here). 
Recall that the equations of general relativity are a set of coupled non-linear
second-order PDEs. Black hole solutions usually depend on the values of the various fields at the 
asymptotic boundary of spacetime as well as the conserved charges carried by the black hole (which 
play the role of integration constants). 
The near-horizon geometry of supersymmetric black holes, by contrast, is governed by the fixed points 
of a set of first-order differential equations, and is completely fixed by the charges of the black hole. 
This is the \emph{black hole attractor mechanism} of~\cite{Ferrara:1995ih}.

The leading-order quantum correction comes from summing the virtual loops of all the quantum excitations 
(graviton, photons,...) in the~AdS$_2 \times$~S$^2$ background, 
using a non-trivial adaptation of standard quantum field theory methods in~\cite{Banerjee:2010qc}. 
The result agrees precisely with the first correction to the leading order result in the asymptotic expansion of the 
Bessel function in~\eqref{RadexpC}:
\be \label{leadinglog}
\log \, d_\text{micro}(N) \= \pi \sqrt{N} - 2 \log N + \text{O} \bigl(1/\sqrt{N} \bigr) \,.
\ee
The next corrections to~$d_\text{micro}$ coming from~\eqref{RadexpC} are in the form of negative powers of~$N$.   
The physical origin of these terms are 
corrections to the local effective action of two-derivative supergravity~\cite{LopesCardoso:1998tkj} 
arising from integrating out the massive modes of the string theory in the background 
of the black hole. They affect the BH entropy in two ways: firstly, in the presence of 
higher-derivative operators in the gravitational effective action the Bekenstein-Hawking formula 
needs to be replaced by the Wald entropy formula~\cite{Wald:1993nt,Iyer:1994ys} and, 
secondly, the black hole solution itself gets corrected by the higher-order effects. 
The reader is referred to~\cite{Sen:2007qy} for a nice review of these ideas and calculations.
The summary is that all such quantum-gravitational calculations in string theory agree with the corresponding 
microscopic results! 

Continuing on to a better approximation to the integer degeneracy, 
one considers the first ($c=1$) term in the Rademacher expansion~\eqref{RadexpC}, i.e.,
\be \label{dFirstBessel}
d_\text{micro}(N)\= \frac{\pi^{9/2}}{8} \, \wt I_{7/2} \bigl(\pi \sqrt{N} \bigr) + \text{O} \bigl(\rme^{\sqrt{N}/2} \bigr) \,.
\ee
This result is interpreted in gravity as the result of summing up the entire perturbation series for the 
quantum entropy of the~$\frac18$-BPS BH.
Such an interpretation is made possible in the framework of the quantum black hole entropy 
expressed as a functional integral over the fluctuating fields 
of supergravity around the near-horizon AdS$_2 \times$~S$^2$ background~\cite{Sen:2008yk,Sen:2008vm}. 
The calculation of this functional integral to all orders in perturbation theory is performed by using 
the technique of supersymmetric localization applied to the gravitational path 
integral \cite{Dabholkar:2010uh,Dabholkar:2011ec, deWit:2018dix, Jeon:2018kec}. 
This reduces the integral over the infinite-dimensional field space to (in this case) a one-dimensional
integral which, with a reasonable assumption about the measure,  
agrees with the representation~\eqref{dFirstBessel} of the~$I$-Bessel function.

Now that the asymptotic series coming from perturbation theory has been summed up into the Bessel function, 
one is now in a position to rigorously discuss the exponentially suppressed corrections coming from the 
terms~$c=2,3,\dots$ in~\eqref{RadexpC}. It was shown in~\cite{Dabholkar:2014ema} that the~$c^\text{th}$
term has the same form as the contribution of a~$\IZ/c\IZ$ orbifold in string theory which contribute to 
the functional integral around asymptotic AdS$_2 \times$~S$^2$.  The Kloosterman sum~$K_c(N)$ is precisely 
reproduced by the functional integral of a certain Chern-Simons theory which emerges in this background.  

Finally,~\cite{Iliesiu:2022kny} revisited and clarified various aspects of the calculation of the supergravity
path integral using supersymmetric localization, 
and applied it to the fluctuations of the gravitational fields in the background of the~$\IZ_c$ orbifold. 
It is important in this calculation to take into account the subtle effects
of a certain mode of the gravitational called the Schwarzian mode\footnote{See~\cite{Mertens:2022irh} for a review
of recent developments including references, and~\cite{Heydeman:2020hhw} for the details of the supersymmetric 
version of the Schwarzian relevant for the discussion in this article.}.
This gives a contribution~$1/c$ to the functional integral (and, therefore, does not affect the~$c=1$ answer). 
Upon putting everything together, one obtains precisely the Bessel function~$\wt I_{7/2}(\pi \sqrt{N/c})$. 
In this manner, the physical quantum-gravitational calculation reproduces, term-by-term, 
the Hardy-Ramanujan-Rademacher expansion.

\section{Mock modular forms \label{sec:MMF}}

The results described in the previous section show that the modular symmetry of the microscopic partition function 
acts as a powerful guiding principle for the quantum-gravitational corrections to the macroscopic black hole. 
This power is based on the agreement of the microscopic and macroscopic pictures, which in turn relies on the 
invariance of the supersymmetric index under a change of parameters of the theory as explained earlier.
Thus one has a perfect one-to-one correspondence between the microstates of the strings and 
branes, and those of the black hole. 

However, there is a subtlety in the statement of invariance of the index. Recall that the main argument 
underlying its invariance is that a change of parameters shifts the energy levels of the 
supersymmetric theory, but states enter or exit the BPS spectrum in boson-fermion pairs, thus not affecting the 
index~$n^B-n^F$. It turns out that there are special codimension-one surfaces
of the parameter space of the theory, called \emph{walls}, upon crossing which 
new BPS solutions enter the spectrum of the theory, which could be purely bosonic or fermionic. 
Typically, the corresponding field configurations are normalizable states in the Hilbert space on one side of the 
wall and become non-normalizable on the other side. 
The index therefore jumps at these walls and this phenomenon is called \emph{wall-crossing}.

The wall-crossing phenomenon manifests itself in supergravity as multi-black hole bound-state configurations 
carrying the same total charges as the original black hole (Fig.~2), that appear on moving across certain 
co-dimension-one surfaces (walls) in the gravitational parameter space~\cite{Denef:2007vg,Sen:2007qy}.  
Our desired single-black-hole states are therefore only part of the full partition function,
and therefore are not, a priori, expected to preserve the symmetries of the full theory. 
For the highly symmetric situation of black holes in~$T^6$ compactifications of string theory  
discussed earlier, 
it was shown in~\cite{Dabholkar:2009dq} that BH bound state solutions do not contribute to the 
relevant supersymmetric index, but in general the wall-crossing phenomenon breaks the 
modular symmetry of the theory. This seems like a potential disaster for the guiding principle! 
\begin{figure}\centering
\includegraphics[width=12cm]{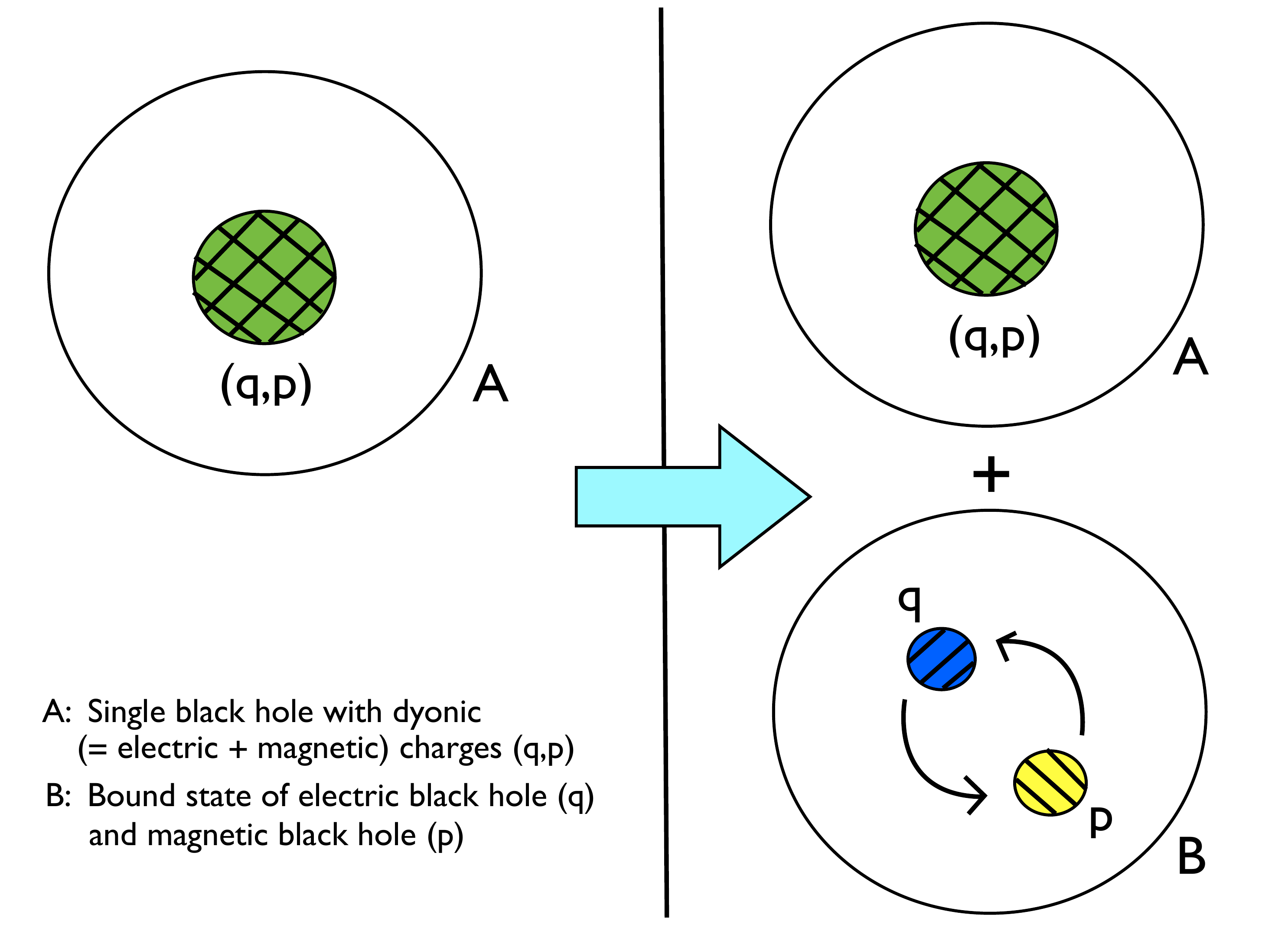}
  \caption{Wall-crossing in supergravity. The vertical line denotes a co-dimension one wall in the parameter space
  of solutions. On the left side of the wall the only solution is the single black hole. 
  On the right side there is an additional solution which is a bound state of two black holes, which is a stationary solution 
  of supergravity with angular momentum carried by electromagnetic fields. 
}
  \label{2BHbndstate}
\end{figure}

The natural question is: can one calculate the supersymmetric index corresponding to the single BH? Does that 
have any interesting modular-like property? 
In general, this is a very difficult question to answer, but in the context of the next simplest theory called 
$\CN=4$ supergravity (coming from string compactifications on~$K3 \times T^2$), 
one has a complete solution to this problem (see the review~\cite{Sen:2007qy}). 
In this case one can show that there is only a specific type of 2-center BH bound state that 
contributes to the index~\cite{Dabholkar:2009dq}. In fact, at any point in the parameter space of string theory, 
the total index can be written as 
\be
d_\text{micro}(n) \= d_\text{1-BH}(n) + d_\text{2-BH}(n)
\ee
where one can explicitly calculate the three components of this equation separately. 

The remarkable fact found in~\cite{Dabholkar:2012nd} is that~$d_\text{1-BH}(n)$ is the Fourier coefficient 
of a \emph{mock modular form}. 
What this means is that one can add a specific correction term (called the \emph{shadow}) to recover 
modular symmetry, analogous to quantum anomalies in physics. 
Examples of such functions (without a definition!) were first discovered by Ramanujan in his famous last letter to 
Hardy, and S.~Zwegers in his PhD thesis~\cite{zwegers2008mock} gave a definition and explained the structure
hidden in Ramanujan's examples (see~\cite{ZagBourbaki,MaassFormsBook}). 

The relations of black hole partition functions and mock modular forms 
were formalized and generalized in~\cite{Dabholkar:2012nd} into a set of theorems 
that apply to a very large class of functions called meromorphic Jacobi forms 
(one sub-family of which is given by the black hole partition functions of $\CN=4$ string theory). 
These theorems have been used to provide exact analytic formulas for the black hole degeneracies extending 
the Hardy-Ramanujan-Rademacher expansion
 \cite{Bringmann:2011, Murthy:2015zzy, Ferrari:2017msn, Chowdhury:2019mnb, 
 LopesCardoso:2020pmp, Cardoso:2021gfg}, 
and to 
prove and analyze conjectures about the positivity of black hole microstate degeneracies 
\cite{Sen:2011ktd, Bringmann:2012zr, Chattopadhyaya:2018xvg, Chattopadhyaya:2020yzl, Govindarajan:2022nzb}. 

These theorems also led to the construction of an infinite number of mock modular forms in~\cite{Dabholkar:2012nd}, 
generalizing the known special examples developed in mathematics starting from the $q$-series of Ramanujan.
The above developments also played a role in the remarkable discoveries in~\cite{Cheng:2012tq}
of \emph{Umbral moonshine} relations between mock modular forms and discrete groups.

\section{Conclusion, and broader relations of modular forms and string theory \label{sec:Conculsions}}

We have described one small corner of the relations of modular forms and modular symmetry 
to the physics of black holes in string theory, and discussed how they (i) provide a guiding principle for the 
quantum-gravitational effects in black holes, (ii) are a powerful tool for calculating black hole entropy, and 
(iii) have fed back and led to new developments in mathematics. 

There are many other very interesting relations of modular forms and string theory (and, more generally, physics) that 
could not be covered in this brief article. 
In particular, modular forms have made an appearance in string theory from the very early days through the 
world-sheet torus amplitudes of free strings, and to estimate the 
Hagedorn growth of states of strings (see~\cite{Green:1987sp}).  
Even before that, the simplest manifestation of modular invariance had appeared in the analysis
of 2d CFT in~\cite{Cardy:1986ie} and led to the famous Cardy formula. 

In the 35 years since then, modular, and more generally automorphic forms, have made multiple appearances 
in string theory (see~\cite{DHoker:2022dxx} for a recent summary). They have been used as a powerful tool 
e.g.~in the description of string worldsheet amplitudes~(see~\cite{Kleinschmidt}). 
The corresponding symmetry has also played the role of a fundamental principle as dualities e.g.~the $SL(2,\IZ)$
duality of supersymmetric Yang-Mills theory (generalizing Montonen-Olive duality) in~\cite{Vafa:1994tf}, 
and of type-IIB string theory in~\cite{Sen:1994fa}, which has given rise to strong constraints on the structure of the 
effective action of string and M-theory~(see~\cite{Green:2005ba}, \cite{Green:2010kv}, \cite{Green:2010wi}, 
\cite{Pioline:2010kb},  \cite{Bossard:2020xod}, \cite{Liu:2022bfg}, and references therein).

As discussed above, black hole microstate degeneracies 
are captured precisely by a modular form in~$T^6$ compactifications of string theory ($\CN=8$ supergravity) 
and by mock modular forms in~$K3 \times T^2$ compactifications ($\CN=4$ supergravity). 
Modular and Jacobi structures in more general Calabi-Yau 
compactifications ($\CN=2$ supergravity) have been discussed in \cite{Gaiotto:2006wm}, \cite{Klemm:2012sx},
\cite{Huang:2015sta}, \cite{Haghighat:2015ega}, \cite{Bouchard:2016lfg}. 
Relations between black hole partition functions, functions called indefinite theta series 
(which are closely related to mock modular forms), and mock modular forms of ``higher depth"  
 have been found in \cite{Manschot:2010sxc}, \cite{Manschot:2010qz}, 
\cite{Alexandrov:2012au}, \cite{Alexandrov:2018lgp}. 

Mock modular forms have also made other appearances in conformal field theory and string theory,
seemingly unrelated to black holes. 
This includes the early work on~$\mathcal{N}=4$ superconformal characters~\cite{Eguchi:1987wf},
Mathieu moonshine~\cite{Eguchi:2010ej} and Umbral moonshine and Niemeier 
lattices~\cite{Cheng:2012tq,Cheng:2013wca}, and relations to K3 surfaces~\cite{Gaberdiel:2011fg, Taormina:2013jza}. 
Another important physical situation where mock modular forms have made an appearance is in the context of 
the completion of the partition function of Vafa-Witten theory~\cite{Dabholkar:2020fde} , and more generally 
four-dimensional~$\CN=2$ gauge theories~\cite{Korpas:2019cwg, Manschot:2021qqe, Aspman:2021kfp, Korpas:2022tij}. 

Physically, the appearance of mock modular behavior is related to the non-compactness of field space,
and consequent holomorphic anomalies. 
This has been studied in the context of supersymmetric quantum mechanics with non-compact target 
\cite{Murthy:2018bzs, Dabholkar:2019nnc}, 
non-compact sigma models and CFT \cite{Eguchi:2010cb,Troost:2010ud,Ashok:2011cy},
\cite{Murthy:2013mya,Ashok:2013pya,Harvey:2014nha,Gupta:2017bcp,Nazaroglu:2016lmr,KumarGupta:2018rac}, 
and string theory \cite{Harvey:2013mda,Cheng:2014zpa,Harvey:2014cva}, and 
AdS/CFT~\cite{Manschot:2007ha}.

More broadly, there are other extremely interesting topics at the modular-forms-physics interface, 
which are slowly being uncovered. 
These include the discussion of black holes and class groups~\cite{Benjamin:2018mdo}, 
the relation of quantum modular forms and 3-manifold invariants \cite{Gukov:2017kmk, Cheng:2018vpl, Garoufalidis:2021lcp}, 
modularity of periods at special points 
of Calabi-Yau moduli spaces~\cite{Moore:1998pn,Candelas:2019llw},
and probably many other treasures that are waiting to be discovered! 

\bibliographystyle{apacite}
%

\end{document}